\documentclass{article}

\usepackage{PRIMEarxiv}

\usepackage[utf8]{inputenc} 
\usepackage[T1]{fontenc}    
\usepackage{hyperref}       
\usepackage{url}            
\usepackage{booktabs}       
\usepackage{amsfonts}       
\usepackage{nicefrac}       
\usepackage{microtype}      
\usepackage{lipsum}
\usepackage{algorithmic}
\usepackage{cite}
\usepackage{textcomp}
\usepackage{amsmath,amssymb,amsfonts}
\usepackage{tabularray} 
\usepackage{array}
\usepackage{color}
\usepackage{subcaption}
\usepackage{multirow}
\usepackage{fancyhdr}       
\usepackage{graphicx}       
\graphicspath{{media/}}     

\title{A Deep-Learning-Based Label-Free No-Reference Image Quality Assessment Metric: Application in Sodium MRI Denoising
}

\author{
  Shuaiyu Yuan\textsuperscript{1*}, Tristan Whitmarsh\textsuperscript{1} ,  Dimitri A Kessler\textsuperscript{1,2},  Otso Arponen\textsuperscript{1},  Mary A McLean\textsuperscript{1},\\ \textbf{Gabrielle Baxter}\textsuperscript{1}, \textbf{Frank Riemer}\textsuperscript{3}, \textbf{Aneurin J Kennerley}\textsuperscript{4,5}, \textbf{William J Brackenbury}\textsuperscript{4}, \\\textbf{Fiona J Gilbert}\textsuperscript{1} \textbf{and Joshua D Kaggie}\textsuperscript{1*} \\
  \textsuperscript{1}University of Cambridge, \textsuperscript{2}Universitat de Barcelona, \textsuperscript{3}Haukeland University Hospital,\\
  \textsuperscript{4}University of York,  \textsuperscript{5}Manchester Metropolitan University\\
  \textsuperscript{*}\texttt{\{sy442, jk636\}@cam.ac.uk} \\
}

\begin{document}
\maketitle

\begin{abstract}
New multinuclear MRI techniques, such as sodium MRI, generally suffer from low image quality due to an inherently low signal. Postprocessing methods, such as image denoising, have been developed for image enhancement. However, the assessment of these enhanced images is challenging especially considering when there is a lack of high resolution and high signal images as reference, such as in sodium MRI. No-reference Image Quality Assessment (NR-IQA) metrics are approaches to solve this problem. Existing learning-based NR-IQA metrics rely on labels derived from subjective human opinions or metrics like Signal-to-Noise Ratio (SNR), which are either time-consuming or lack accurate ground truths, resulting in unreliable assessment. We note that deep learning (DL) models have a unique characteristic in that they are specialized to a characteristic training set, meaning that deviations between the input testing data from the training data will reduce prediction accuracy. Therefore, we propose a novel DL-based NR-IQA metric, the Model Specialization Metric (MSM), which does not depend on ground-truth images or labels. MSM measures the difference between the input image and the model’s prediction for evaluating the quality of the input image. Experiments conducted on both simulated distorted proton T1-weighted MR images and denoised sodium MR images demonstrate that MSM exhibits a superior evaluation performance on various simulated noises and distortions. MSM also has a substantial  agreement with the expert evaluations, achieving an averaged Cohen’s Kappa ($\kappa$) coefficient of 0.6528, outperforming the existing NR-IQA metrics.
\end{abstract}

\keywords{No-reference Image Quality Assessment \and Sodium MRI \and Deep learning \and Model Specialization}

\section{Introduction}
{M}{AGNETIC} Resonance Imaging (MRI) is a powerful and versatile imaging technique widely used in clinical diagnosis and research. Traditional MRI will mainly detect the distribution and properties of hydrogen nuclei, namely protons, due to their abundance in water and fat, providing high-quality anatomical and compositional characteristics of the human body. Sodium MRI research probes sodium ionic concentrations with molecular specificity that is unobtainable with proton MRI \cite{b1}. Sodium ionic concentrations maintain homeostasis from physiological processes that include cellular pumps such as Na\textsuperscript{+}/K\textsuperscript{+} -ATPase , and from cellular membrane integrity \cite{b2}. Disruption of normal physiological processes will cause ionic imbalances, which could indicate disease states \cite{b2}. The challenge faced with sodium MRI is its low signal-to-noise ratio (SNR). Compared to hydrogen, in vivo sodium concentrations are approximately a factor of 1000 lower \cite{b3}. A fundamental physical property of a nucleus is its gyromagnetic ratio, which relates to the proportion of nuclei polarized, and which for sodium is 26\% that of proton. The lower polarization of sodium reduces signal further by a factor of ~1/16  \cite{b4}. From these factors alone, the signal of sodium can be 16,000 times lower than that of hydrogen during conventional MRI. This significantly lower SNR of sodium reduces its potential in diagnostic practice.

Sodium MRI image quality has been constantly made more feasible through improved hardware, acquisition, reconstruction, and post-processing techniques. For example, MRI scanners with high field strengths polarize the nuclei and thus improve the signal proportionally. Acquisition techniques such as ultra-short echo time (UTE) sequences improve the signal acquisition of sodium MRI before rapid and significant signal decay \cite{b5}. Post-processing reconstruction algorithms have improved the image quality and enable new acquisition schemes, which rely on non-Cartesian reconstruction techniques, such as the gridding algorithms\cite{b6}, non-uniform fast Fourier transform (NUFFT) \cite{b7} or iterative reconstruction \cite{b8}. Deep learning algorithms have also demonstrated further image improvements with image denoising and super-resolution methods \cite{b9,b10}. 

Reliable Image Quality Assessment (IQA) metrics are required to assess the quality of enhanced images regardless of the enhancement approach taken. IQA metrics can be divided into two classes: full-reference IQA (FR-IQA) and no-reference IQA (NR-IQA) \cite{b11}. Medical imaging commonly uses FR-IQA metrics to assess methodological improvements, such as Peak Signal-to-Noise Ratio (PSNR), Structural Similarity (SSIM) \cite{b11} and Feature Similarity (FSIM) \cite{b12}. However, FR-IQA metrics are not able to evaluate the quality of sodium MR images that have been digitally enhanced such as through image denoising methods. High-quality reference images are difficult, if not impossible, to obtain for sodium MRI in patients as a result of requiring intolerable scan times and physics limitations from limited SNR, as well as rapid signal decays during acquisition and image blurring.

For this reason, NR-IQA metrics are required for the evaluation of sodium MRI denoising. An example of a NR-IQA metric is the commonly used Blind/Referenceless Image Spatial Quality Evaluator (BRISQUE) \cite{b13}. BRISQUE extracts up to 36 features, including shape, mean and variance of different parts of the image. These 36 features are fed into a regressor, such as a Support Vector Machine (SVM), which then generates a final quality score. With the development of DL algorithms, several DL-based NR-IQA metrics have been proposed \cite{b14}. The common pipeline of a DL-based NR-IQA metric is the use of a pre-trained DL model, such as VGG16 \cite{b15}, ResNet18 \cite{b16} or ViT \cite{b17}, to extract numerous high-level image features, which features are subsequently fed into a regression or classification model for evaluating the image quality. NR-IQA can also be performed with a fusion of several DL models \cite{b18,b19}. Labels are used to train DL-based NR-IQA metrics, which labels can include SNR \cite{b20}, error maps \cite{b21}, synthetic scores derived from FR-IQA metrics \cite{b22}, and metrics based on intermediate layers of deep learning models \cite{b23}.

In the case of evaluating the quality of denoising enhancements of sodium MR images, these commonly used training labels are insufficient. SNR alone does not accurately capture improvements in image quality, as image blurring and feature loss is frequently associated with improved SNR. Alternative training labels, such as mean squared error, are infeasible since they rely on high-quality reference images, which are impractical to be obtained for sodium MRI that lack large ground-truth datasets, especially in the case of patient data. An alternate labelling method is through human evaluations by expert readers to provide a Mean Opinion Score (MOS) of the image quality \cite{b24}. While MOS is commonly considered the gold standard of image quality, the primary drawback of MOS to assess denoised sodium MR images is that an expert reader has biases and may value different aspects of an ‘improved’ image, such as blurring that improves the overall signal but reduces visible features. The reliance on human evaluations also makes obtaining MOS a time-consuming and expensive process, limiting the numbers of images that can be assessed.

This paper introduces a novel, label-free DL-based NR-IQA metric, the \textbf{M}odel \textbf{S}pecialization \textbf{M}etric (MSM), to address the limitations in existing IQA metrics for evaluating the quality of denoised sodium MR images. This metric is designed to obviate the need for human assessment and avoid reliance on objective labels that may be impractical to obtain given the specific constraints of sodium MRI. The metric is based on the observed phenomenon that the quality of the DL model’s prediction decreases when the input testing data distribution is different from the model’s training data distribution. Since such discrepancies are associated with the input image quality, we propose that the measurement of distance or similarity between the input image and corresponding model’s prediction can be used to assess the image quality. The main contributions of this study are as follows:

1) A DL-based NR-IQA metric termed MSM is developed to assess image quality when lacking high-quality reference images. MSM avoids the reliance on labels for training, which can be impractical to obtain.

2) Various experiments and comparisons with other NR-IQA and FR-IQA metrics are presented and demonstrate the superior performance of MSM to assess image quality.

\section{Method}
In this section, we: A) introduce the concept of Model Specialization, illustrating how deviations from training data impact model performance; B) develop the Model Specialization Metric (MSM), a deep learning-based NR-IQA metric designed to assess image quality without the reliance of any label for training; C) Evaluate two models to serve as the backbone for MSM, namely, U-net, a CNN-based model, and SwinIR, a Transformer-based model; D) outline the data preparation strategies, which include using images with predetermined noise levels and distortions, images from a generative model, and sodium MRI denoising; E) describe the evaluation metrics used to validate MSM, including Pearson Linear Correlation Coefficient (PLCC), Spearman Rank Correlation Coefficient (SRCC), and Cohen's Kappa ($\kappa$) Coefficient.

\begin{figure}[!t]
\centerline{\includegraphics[width=\columnwidth]{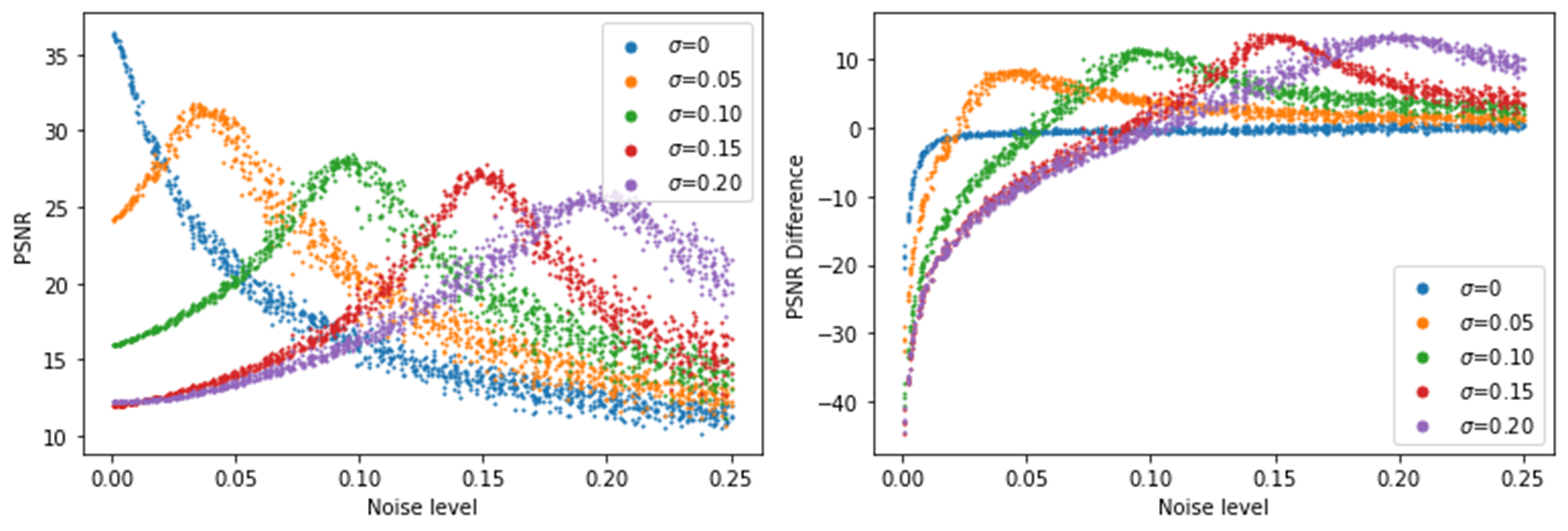}}
\caption{(Left) An example of Model Specialization. A reference image was distorted using Gaussian noises with varying noise levels from 0 to 0.25. The graph plots the PSNR values of U-net denoised results against the noise levels of the corresponding noisy images. (Right) A graph showing the association between the noise levels of input image and the PSNR value difference between the input and the denoised images. The legend $\sigma$ indicates the noise levels at which the U-net was trained, specifically 0, 0.05, 0.10, 0.15, and 0.20.}
\label{fig1}
\end{figure}

\subsection{Model Specialization}
In the context of this study, we introduce and define the concept of ‘\textbf{Model Specialization}’. This concept is that when a DL model is trained on a specific dataset tailored for a particular application, the model becomes highly specialized. Consequently, any deviation in the input testing data from the training data will decrease the model’s predictive performance.

Fig. 1 (left) illustrates the concept of Model Specialization. A denoising model, U-net \cite{b25}, is trained on 24 reference images from the TID2013 dataset \cite{b26} with additional fixed Gaussian noises with 0 means and four different noise levels ($\sigma$) from 0 to 0.20, respectively. A separate reference image with varying noise levels is denoised by this trained U-net. The PSNR values of denoised results are shown in Fig.1 (left). The figure indicates that when the DL model is trained with a specific noise level, the model yields optimal denoising results only when applied to noisy images that have the same noise level as the training noise. Notably, the model's denoising performance decreases as the noise level of the input images deviates from that of the training data, irrespective of whether the noisy images exhibit lower or higher noise levels. Similar phenomena can be observed with various distortions and different noise types, such as Rician noise. The principle of Model Specialization serves as the core concept of our novel label-free NR-IQA metric.

\subsection{Model Specialization Metric}
In Fig. 1 (right), the PSNR values of the corresponding noisy images and the differences between the noisy and denoised images are plotted. The results indicate that the discrepancy between the input image and the model’s prediction is related to the quality of the input image. Notably, when the model is trained on noise-free images ($\sigma$=0), this relationship becomes monotonic. As a result, in this case the measurement measuring the discrepancy using the distance or similarity  between the input image and the model’s prediction can serve as a metric, termed Model Specialization Metric (MSM), to assess the quality of the input image’s quality.

MSM is defined as:
\begin{equation}\text{MSM} = \text{difference}(I, M(I))\nonumber\end{equation}
Where $I$ is the input image, $M$ is a pre-trained DL model used as a backbone model in this metric and $M(I)$ is the model’s predicted output. The difference can be measured with distances, such as L1 distance and L2 distance, or through a similarity, such as PSNR and SSIM. To avoid confusion with IQA metrics such as PSNR and SSIM, the similarity measurement is denoted as $S_{PSNR}$ and $S_{SSIM}$. MSM aims to measure the difference between the $I$ and $M(I)$, and use this difference as the assessment of the image quality of $I$. 

A Taylor Series expansion is used to explain why the difference between the $I$ and $M(I)$ can be used to assess the image quality. The Taylor series expansion of a function $f(x)$ around a point $a$ is given by:
\begin{equation}f(x) = \sum_{n=0}^{\infty} \frac{f^{(n)}(a)}{n!} (x - a)^n\label{eq}\end{equation}
Considering the DL model $M$ as a function of the  $I$, expanded around the ground truth ($Gt$), and ignoring the second and higher-order terms, get results in:
\begin{equation}M(I) = M(Gt) + M'(Gt)(I - Gt)
\label{eq}
\end{equation}
Then, if $M$ is trained to map the ground truth to itself, when the loss converges, $M'(Gt)$ can be assumed to be a constant  $\frac{1}{\alpha}$:
\begin{equation}
M(I) = M(Gt) + \frac{1}{\alpha}(I - Gt)
\label{eq}
\end{equation}
 $M(Gt)$ can be approximated with $M(Gt)=Gt$ to obtain:
\begin{equation}
I - Gt = \frac{\alpha}{\alpha-1} (1 - M(I))
\label{eq}
\end{equation}
Equation (4) indicates that the difference between the input image and the ground truth is proportional to the difference between the input and the model’s predicted output.  An important precondition must be satisfied to achieve (4): the backbone model should be trained to map the ground truth images back onto themselves. 

In the case of this paper, the model is trained to map the undistorted images to themselves, which means only clean images are required. This training strategy eliminates the need for labelling or the introduction of synthetic distortions during training, significantly reducing the complexity and resources required for dataset preparation. As a result, the proposed MSM model can assess images with a wide range of distortion types without retraining or adaptation to each new distortion category.

\subsubsection{Backbone Model}
This study evaluates two distinct backbone models to implement the MSM: the U-net, a classic CNN-based model and SwinIR \cite{b27}, a Transformer-based model. MSM implemented with these two backbone models are referred to as $MSM_{Unet}$ and $MSM_{SwinIR}$, respectively. 

This approach allows us to assess the efficacy of the MSM method across different backbone models, providing insight into how each model's inherent strengths and weaknesses influence the MSM’s performance.

\subsection{Data Preparation}
We employ three approaches to prepare the data to test the evaluation performance of the MSM, which include simulating noises and distortions, utilizing generative models, and image denoising. 
\begin{figure}[!t]
\centerline{\includegraphics[width=0.75\columnwidth]{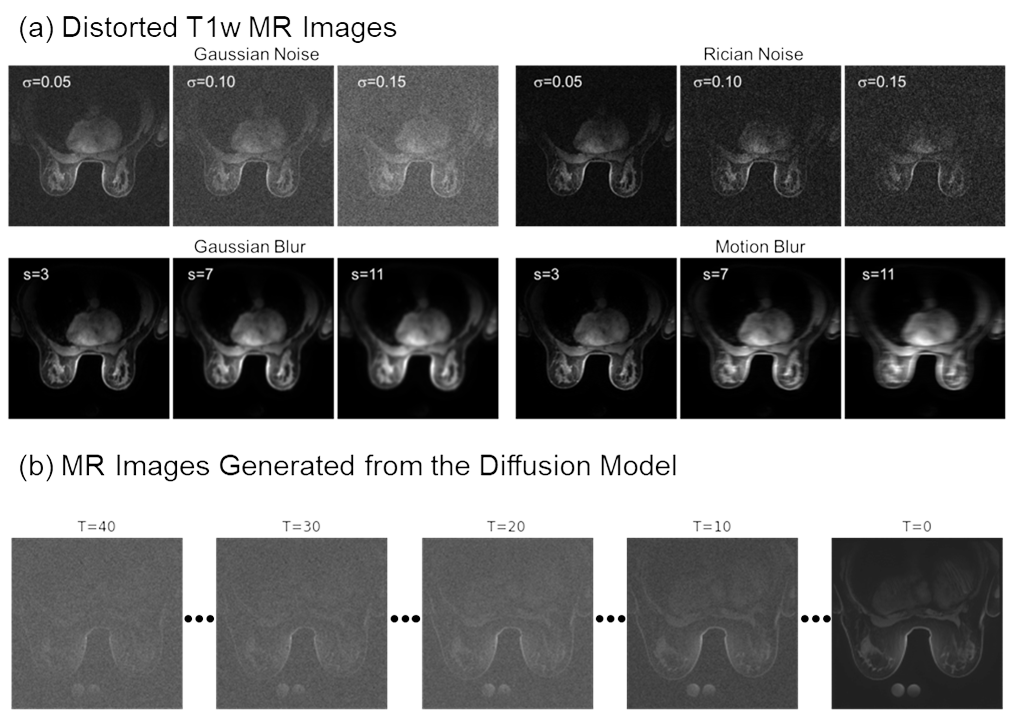}}
\caption{Examples of (a) proton T$_1$w MR image with additional Gaussian and Rician noise, with standard deviations of 0.05, 0.10 and 0.15, and blurred using a Gaussian and a motion blur filter with filter sizes of 3, 7 and 11; (b) Synthetic MR images generated from DDPM at timesteps of 40, 30, 20, 10 and 0.}
\label{fig2}
\end{figure}

\subsubsection{Noise and Distortion Simulation}
Noise and blurriness are two main types of image losses  that exist in MRI, especially non-proton MR images. \cite{b28} As shown in Fig. 2(a), two noise types, Gaussian noise and Rician noise, and two blurriness types, Gaussian blur and Motion blur, are simulated in this paper. The two types of noise, Gaussian and Rician noise, are simulated with a random standard deviation ($\sigma$) from 0 to 0.25 of the maximum image intensity and added to the high-quality proton T1-weighted (T$_1$w) MR images. Regarding blurriness, a Gaussian blur and a horizontal motion blur filter with a random size (s) from 3 to 51 pixels is applied to the T$_1$w MR images using OpenCV \cite{b29}. Even though there is not a strictly definable metric for absolute image quality score for many types of image distortions, increases in the noise standard deviation and blur filter kernel size are assumed to be directly associated with worse image quality. Noise and distortion simulations are then used to measure the alignment of IQA metrics’ predictions for their ability to evaluate image quality.

\subsubsection{MR Images Generated from the Diffusion Model}
While directly adding Gaussian or Rician noise is straightforward, it means that the pattern of noise is purely random and does not consider image content or contextual information. To generate a more complex noise model which is less uniform, this paper also generates noisy images using a diffusion model, Denoising Diffusion Probabilistic Models (DDPM) \cite{b30}.

In a DDPM, the noise added at each timestep is typically conditional on the current state of the image. This means that the noise can adapt based on the evolving features of the image through the diffusion process. This could result in noise patterns that are less predictable and possibly more akin to real-world noise, such as may occur with imperfections in magnetic field gradients. Fig. 2(b) gives an example of the synthetic noisy MR images generated from DDPM. $T$ means the timestep at which the denoising process stops before reaching a clean image. Lager $T$ correspond to larger noise levels and worsening image quality. The IQA metrics’ evaluation performance can be assessed based on the correlation between the evaluation results and the timesteps.

\subsubsection{Sodium MRI Denoising}
Two non-DL denoising methods, median filter and Block Matching 3D (BM3D) \cite{b31}, and three supervised CNN-based denoising models, U-net, DnCNN \cite{b32} and ResUnet \cite{b33}, are used to denoise sodium MR images. An example of the denoised results is shown in Fig.3.

When denoising sodium MR images with supervised DL models, due to the unavailability of ground-truth references for sodium MR images that can match the resolution and SNR of proton T1-weighted MRI, synthetic sodium MR images are created for training. A synthetic sodium MR image is formed by introducing noise into a proton T$_1$w MR image using the equation,
\begin{equation}R = \sqrt{\left(S + \frac{N}{\sqrt{2}}\right)^2 + \left(\frac{N}{\sqrt{2}}\right)^2}\nonumber\end{equation}
Where $R$ is the synthetic sodium MR image, $S$ is the T$_1$w MR image and $N$ is the noise obtained from native sodium MRI data in an distinct patient, which noise is obtained from the slices of the native sodium MRI data that contains no signal.

\subsection{Implementation}
All networks were implemented using PyTorch and trained and tested on an Nvidia A100 GPU. During the data preparation process, the DDPM had 1000 diffusion timesteps, using a linear noise schedule and was trained with a learning rate of 0.0002, batch size of 10 for 1000 epochs. The DL models for sodium MR image denoising were trained using the Adam optimizer, with a batch size of 10 for 250 epochs and a learning rate of 0.0001 using an MSE loss function.

Regarding backbone models of MSM, the U-net utilized three downsampling layers and 32 output channels in the initial downsampling layer with an expansion factor of two. Owing to GPU memory limitations, the SwinIR model utilized 4 attention heads in each Swin transformer block. The backbone models were trained using the Adam optimizer, batch size of 10 for 200 epochs and a learning rate of 0.001 that exponentially decayed with a factor of 0.99. Finally, the learning-based NR-IQA metrics were fine-tuned using a learning rate of 0.0005 for 100 epochs.

\subsection{Evaluation}
\subsubsection{Distorted T$_1$w MR Images and Generated MR Images}
The evaluation results of IQA metrics are assessed using two correlation coefficients: Pearson Linear Correlation Coefficient (PLCC) and Spearman Rank Correlation Coefficient (SRCC). These metrics are tested with distorted T$_1$w MR images and diffusion DDPM generated MR images. PLCC measures the linear correlation between the evaluation results and the known distortion levels and is calculated as:
\begin{equation}\text{PLCC} = \frac{\sum_{i=1}^n (x_i - \bar{x})(y_i - \bar{y})}{\sqrt{\sum_{i=1}^n (x_i - \bar{x})^2 \sum_{i=1}^n (y_i - \bar{y})^2}}\nonumber\end{equation}
Where $n$ is the testing dataset size, $x_i$ and $y_i$ are the individual sample points of evaluation results and known distortion levels, respectively. $\bar x$  represents the sample mean and analogously for $\bar y$.

SRCC assesses the monotonicity between the evaluation results and the known distortion levels and is given by:
\begin{equation}\text{SRCC} = 1 - \frac{6 \sum_{i=1}^n (x_i - y_i)^2}{n(n-1)}\nonumber\end{equation}
Where $x_i - y_i$ is the difference between each pair of evaluation results and the corresponding distortion levels. For simplicity, all results are presented as absolute values, although negative correlations are possible. A higher magnitude of PLCC and SRCC value means the IQA metric has a better evaluation performance.

\subsubsection{Denoised Sodium MR Images}
Unlike the evaluation of distorted proton T$_1$w MR images and DDPM generated MR images, which have known distortion levels, these denoised sodium MR images do not have objective quality labels. Therefore, three experts in MRI were invited to evaluate the quality of denoised sodium MR images, according to noise level, edge sharpness and feature distortions. Instead of asking the experts to give a specific score to the images, the experts only need to determine which image has higher quality for any pair of denoised images.

We use Cohen's Kappa ($\kappa$) to measure the agreement between experts and the IQA metrics. Cohen's Kappa is commonly used to evaluate the level of agreement between two metrics which classify items into mutually exclusive categories \cite{b34}. The $\kappa$ is given as:
\begin{equation}\kappa = \frac{p_o - p_e}{1 - p_e}\nonumber\end{equation}
Where $p_o$ is the observed agreement proportion and $p_e$ represents the expected agreement proportion by chance, which equals the sum of the products of the probabilities for each metric to choose each category by chance. A $\kappa$ value of 1 indicates perfect agreement, a $\kappa$ value of 0 indicates that the agreement is no better than chance, while a negative $\kappa$ value means the agreement is worse than chance.

\section{Experiments}
\subsection{Dataset}
\subsubsection{Breast MRI Data}
This study used data from a cohort of 27 breast cancer patients. Sodium MRI data was acquired alongside conventional proton T1-weighted MRI data for this study. The data was obtained with ethical approval (IRAS ID: 260281; West Midlands - Black Country Research Ethics Committee) and informed consent. The data was acquired following an MRI protocol formerly published \cite{b35}. Two datasets were built in this study.

The \textbf{T$_1$w MRI Dataset} consisting of 810 slices of proton T$_1$w MR breast images from 27 patients. The T$_1$w MRI Dataset is used to train the MSM model and evaluate the MSM’s performance with simulated distortions.

The \textbf{Sodium MRI Dataset} consisting of 810 slices of sodium MR images from 27 patients. The Sodium MRI Dataset aims to assess the IQA metrics’ performance with unknown distortions and how the evaluation results align with experts’ opinions. 

\subsubsection{Brain MRI Data}
The \textbf{Brain Tumor MRI Dataset} is publicly available \cite{b36}, containing 4787 slices of 109 brain MRI scans. For this study, 500 slices were randomly selected from this dataset and applied with the distortions. The Brain Tumor MRI Dataset is employed to investigate the performance of MSM on an unseen dataset as external validation.

\subsection{Comparison of Different IQA Metrics}
To evaluate the ability of MSM to assess image quality, we compare with six FR-IQA metrics, including PSNR, SSIM, FSIM, Fréchet Inception Distance (FID), Learned Perceptual Image Patch Similarity (LPIPS) \cite{b37} and Deep Image Structure and Texture Similarity (DISTS) \cite{b38}. We also compare MSM with five state-of-the-art NR-IQA metrics, including BRISQUE, Natural Image Quality Evaluator (NIQE) \cite{b39}, From patches to pictures (PaQ-2-PiQ) \cite{b40}, Multi-Scale Image Quality Transformer (MUSIQ) \cite{b41} and Weighted Average Deep Image QuAlity Measure for NR IQA (WaDIQaM) \cite{b42}. These metrics are available in the following repository \cite{b43}. Among these metrics, LPIPS, DISTS, PaQ2PiQ, MUSIQ and WaDIQaM are DL-based IQA metrics, which were initially trained with two large datasets, KADID-10k \cite{b44} and KonIQ-10k \cite{b45}, respectively. For a fair comparison, these metrics are fine-tuned using the same training dataset as MSM, except BRISQUE, LPIPS and DISTS, where the fine-tuning decreases their performance.

All IQA metrics are first trained and assessed on the T$_1$w MRI Dataset with the aforementioned simulations of uniform noise, distortions, and DDPM-noised MR images. Every 3 patients are divided as a group and 9-fold cross validation is conducted to reduce the bias of randomness. To validate the cross-dataset performance of our model, we also use the T$_1$w MRI Dataset for training and the Brain Tumor MRI dataset for testing. NR-IQA metrics are further evaluated on the Sodium MRI Dataset and their evaluation results are assessed based on experts’ opinions.

\subsection{Ablation Study}
A systematic ablation study was performed to assess the performance of two backbone models, U-net and SwinIR, across a variety of configurations. This involved testing each model with three different loss functions, L1 loss, L2 loss, and perceptual loss \cite{b46}, and two distance measurements, L1 distance and L2 distance and two similarity metrics, $S_{PSNR}$ and $S_{SSIM}$, on the T$_1$w MRI Dataset with simulated noises and distortions.  The ablation study was employed to determine which combination of loss function and distance or similarity measurement for the MSM can achieve the best performance.

\section{Results}
\subsection{Sodium MRI Denoising Qualitative Results}
\begin{figure}[!t]
\centerline{\includegraphics[width=0.75\columnwidth]{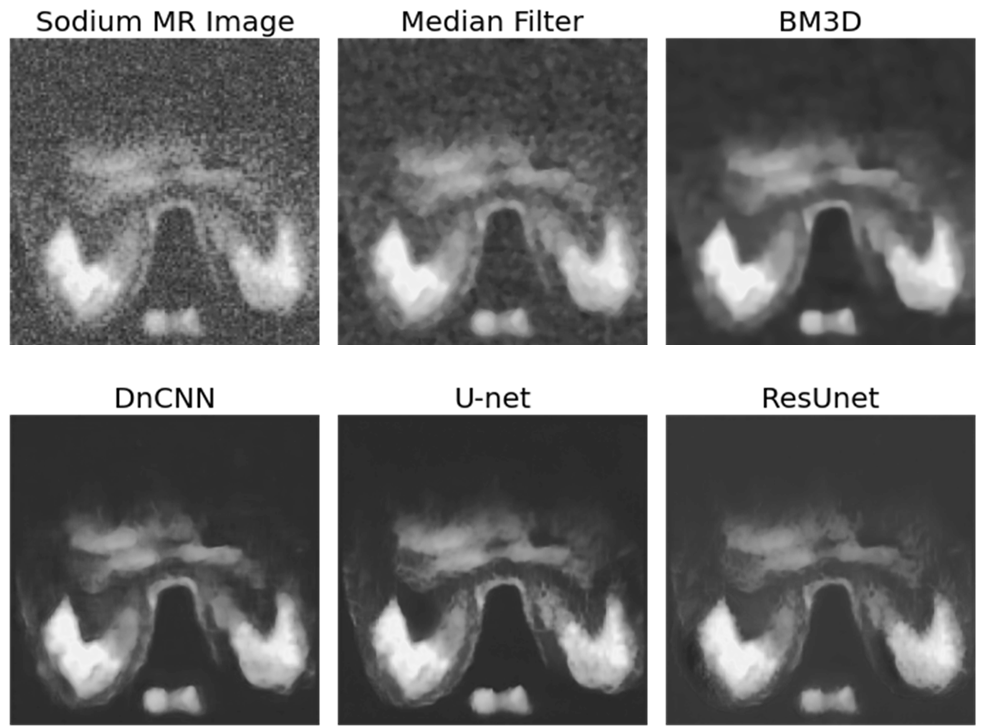}}
\caption{Examples of  an authentic sodium MR image and corresponding denoising results of median filter, BM3D, DnCNN, U-net and ResUnet.}
\label{fig3}
\end{figure}
Fig. 3 presents examples of a slice of sodium MR image and its corresponding denoising results. Specifically,, the median filter fails to remove the noise points. DnCNN and BM3D and DnCNN tend to introduce blurriness to the image, compromising the clarity and overall quality of the images. In contrast, the denoising results of U-net and ResUnet preserve more detailed structures that aligning more closely with the original sodium MR image and indicating better image quality. 

While visual assessment provides an immediate understanding of the denoising effectiveness, it is inherently subjective and can overlook subtle but important differences in image quality. Quantitative assessment is necessary to objectively evaluate the performance of denoising algorithms. However, assessing the quality of denoised sodium MR images quantitatively is complex, as their distortions often involve a combination of noise types and structural variations that cannot be directly quantified using predetermined parameters.

\subsection{Ablation Study Result}
\begin{table*}[ht]
\caption{The SRCC/PLCC results of the ablation study results on the T$_1$w MRI Dataset. The highest average results are highlighted in \textbf{bold}}
\centering
\resizebox{\linewidth}{!}{%
\begin{tblr}{
  cell{2}{1} = {r=3}{},
  cell{5}{1} = {r=3}{},
  hline{1,2,5,8} = {-}{0.12em},
}
Model & Loss       & Distance & Gaussian   Noise & Rician   Noise & DDPM          & Gaussian   Blur & Motion   Blur & Average       \\
\SetCell{}{}{U-net}  & L1 & $S_{PSNR}$ & 0.9715/0.9111 & 0.9654/0.8756 & 0.8010/0.6910 & 0.8291/0.8051 & 0.7750/0.7561 & 0.8701/0.8074 \\
                        & L2 & $S_{PSNR}$ & 0.9683/0.8976 & 0.9659/0.8517 & 0.8100/0.6221 & 0.7596/0.7386 & 0.7722/0.7584 & 0.8552/0.7737 \\
      & Perceptual & L2       & 0.9705/0.9669    & 0.9661/0.8849  & 0.8114/0.7425 & 0.8040/0.7831   & 0.8004/0.7796 & \textbf{0.8705/0.8314} \\
\SetCell{}{}{SwinIR} & L1 & $S_{PSNR}$ & 0.9322/0.8751 & 0.9538/0.9105 & 0.7033/0.5706 & 0.8381/0.8250 & 0.7700/0.7736 & 0.8395/\textbf{0.7930} \\
                        & L2 & $S_{PSNR}$ & 0.9367/0.9105 & 0.9538/0.9477 & 0.7169/0.5727 & 0.7117/0.6202 & 0.6435/0.5423 & 0.7925/0.7187 \\
      & Perceptual & $S_{SSIM}$    & 0.8970/0.8858    & 0.8217/0.8030  & 0.6619/0.6725 & 0.9291/0.7805   & 0.9106/0.8213 & \textbf{0.8441}/0.7926
\end{tblr}
}
\end{table*}

Table 1 lists the best SRCC/PLCC results for each backbone model regarding different loss functions and distance measurements. The best averaged results are emphasized in \textbf{bold}. These results not only validate the effectiveness of the ground-truth-to-ground-truth training strategy across both U-net and SwinIR models but also present that the proposed MSM method can effectively assess the image quality with the simulated noises and distortions.

For the U-net, the results reveal that employing L1 loss in conjunction with $S_{PSNR}$ achieves the highest SRCC/PLCC value of 0.8291/0.8051 when evaluating Gaussian Blur. However, the integration of perceptual loss and L2 distance obtains the highest scores of 0.9705/0.9669, 0.9661/0.8849, 0.8114/0.7425 and 0.8040/0.7796 when assessing the Gaussian noise, Rician noise, DDPM noise and Motion blur distortions, respectively. Therefore, the utility of perceptual loss and L2 distance is recommended when using U-net as the backbone model.

Similarly, in terms of the SwinIR model, when SwinIR is trained using perceptual loss and $S_{SSIM}$, despite its marginally lower performance in evaluating Gaussian, Rician noises and DDPM noise compared to other configurations, it demonstrates markedly superior results for Gaussian blur and motion blur distortions, with the SRCC/PLCC result of 0.9291/0.7805 and 0.9106/0.8213. As a result, when using SwinIR as the backbone model, it is configured with perceptual loss and $S_{SSIM}$.

\subsection{Results of Distorted and Generated MR Images}
\begin{table*}[ht]
\caption{The average SRCC/PLCC results of six FR-IQA and seven NR-IQA metrics on the T$_1$w MRI Dataset. The highest columnar value is highlighted in \textbf{bold} and the second highest SRCC or PLCC value is marked in \underline{\textit{italics and underline}}.}
\centering
\resizebox{\linewidth}{!}{%
\begin{tblr}{
  cell{2}{1} = {r=6}{},
  cell{8}{1} = {r=7}{},
  vline{2,3,8} = {-}{0.12em},
  hline{1,2,8,15} = {-}{0.12em},
}
&                  & Gaussian   Noise & Rician   Noise & DDPM          & Gaussian   Blur & Motion   Blur & Average       \\
\SetCell{}{}{FR-IQA} & PSNR             & 0.9630/0.9080    & 0.9646/0.9041  & 0.7649/0.4948 & 0.8108/0.8028   & 0.7253/0.7120 & 0.8457/0.7643 \\
                        & SSIM             & 0.9548/0.8161    & 0.9405/0.8095  & 0.7479/0.6999 & 0.7753/0.7751   & 0.7181/0.7252 & 0.8273/0.7652 \\
                        & FSIM             & 0.8730/0.8342    & 0.8498/0.8231  & 0.7433/0.7298 & \textbf{0.9820}/\textbf{0.9798}   & \textbf{0.9663}/\textbf{0.9381} & 0.8829/\underline{\textit{0.8610}} \\
                        & FID              & 0.7774/0.8151    & 0.6761/0.7574  & 0.7449/0.7447 & 0.6981/0.7257   & 0.5596/0.6121 & 0.6912/0.7310 \\
                        & LPIPS            & 0.9531/0.8938    & 0.9608/0.9039  & 0.8053/0.7717 & 0.8776/0.8634   & 0.8965/0.8745 & \textbf{0.8987}/\textbf{0.8615} \\
                        & DISTS            & 0.8783/0.8504    & 0.8817/0.8420  & 0.7742/0.7247 & 0.9561/0.9233   & \underline{\textit{0.9403}}/\underline{\textit{0.9040}} & \underline{\textit{0.8861}}/0.8489 \\
\SetCell{}{}{NR-IQA} & BRISQUE          & 0.9648/0.8998    & 0.9566/0.8085  & 0.7561/0.6730 & 0.9541/0.8895   & 0.5534/0.5731 & 0.8370/0.7688 \\
                        & NIQE             & 0.8535/0.8448    & 0.8448/0.8454  & 0.6625/0.6815 & \underline{\textit{0.9599}}/\underline{\textit{0.9469}}   & 0.6451/0.6676 & 0.7932/0.7972 \\
                        & MUSIQ            & 0.9665/\textbf{0.9690}    & 0.9655/\underline{\textit{0.9640}}  & \textbf{0.8386}/\textbf{0.8425} & 0.5399/0.6727   & 0.6962/0.6916 & 0.8013/0.8280 \\
                        & PAQ2PIQ          & 0.9450/0.9262    & 0.9598/0.9438  & 0.8073/0.8145 & 0.7355/0.7212   & 0.7699/0.7095 & 0.8435/0.8231 \\
                        & WaDIQaM          & \underline{\textit{0.9688}}/0.9656    & \textbf{0.9706}/\textbf{0.9641}  & 0.8084/\underline{\textit{0.8197}} & 0.7208/0.6931   & 0.5595/0.6570 & 0.8056/0.8199 \\
                        & $MSM_{Unet}$ (ours)   & \textbf{0.9705}/\underline{\textit{0.9669}}    & \underline{\textit{0.9661}}/0.8849  & \underline{\textit{0.8114}}/0.7425 & 0.8040/0.7831   & 0.8004/0.7796 & 0.8705/0.8314 \\
                        & $MSM_{SwinIR}$ (ours) & 0.8970/0.8858    & 0.8217/0.8030  & 0.6619/0.6725 & 0.9291/0.7805   & 0.9106/0.8213 & 0.8441/0.7926
\end{tblr}
}
\end{table*}
\begin{table*}[ht]
\caption{The average SRCC/PLCC results of six FR-IQA and seven NR-IQA metrics on he Brain Tumor MRI Dataset. The highest columnar value is highlighted in \textbf{bold} and the second highest SRCC or PLCC value is marked in \underline{\textit{italics and underline}}.}
\centering
\resizebox{\linewidth}{!}{%
\begin{tblr}{
  cell{2}{1} = {r=6}{},
  cell{8}{1} = {r=7}{},
  vline{2,3,8} = {-}{0.12em},
  hline{1,2,8,15,17} = {-}{0.12em},
  }
&                        & Gaussian   Noise & Rician   Noise & DDPM          & Gaussian   Blur & Motion   Blur & Average       \\
\SetCell{}{}{FR-IQA} & PSNR                   & 0.9736/0.9115    & 0.9732/0.9056  & 0.7735/0.5408 & 0.7783/0.7602   & 0.8147/0.7642 & 0.8627/0.7765 \\
                        & SSIM                   & 0.9524/0.8926    & 0.9524/0.8885  & \textbf{0.8346}/\textbf{0.7809} & 0.8942/0.8891   & 0.8579/0.8401 & 0.8983/0.8582 \\
                        & FSIM                   & 0.9570/0.9597    & 0.9643/0.9678  & 0.8207/0.6836 & 0.8009/0.8128   & 0.8488/0.8475 & 0.8783/0.8543 \\
                        & FID                    & 0.8297/0.8370    & 0.7638/0.7784  & 0.7321/0.6698 & 0.6988/0.7122   & 0.6119/0.6249 & 0.7273/0.7245 \\
                        & LPIPS                  & 0.9720/0.9623    & \underline{\textit{0.9755}}/\underline{\textit{0.9709}}  & 0.8208/0.7157 & 0.9194/0.8846   & \textbf{0.9336}/\textbf{0.8948} & \underline{\textit{0.9243}}/\textbf{0.8857} \\
                        & DISTS                  & 0.9694/0.9356    & 0.9709/0.9430  & \underline{\textit{0.9267}}/0.6832 & \underline{\textit{0.9724}}/\textbf{0.9578}   & \underline{\textit{0.9200}}/\underline{\textit{0.8821}} & \textbf{0.9319}/\underline{\textit{0.8803}} \\
\SetCell{}{}{NR-IQA} & BRISQUE                & 0.9517/0.8876    & 0.9436/0.8661  & 0.8128/0.7250 & \textbf{0.9821}/0.9436   & 0.3283/0.3956 & 0.8037/0.7636 \\
                        & NIQE                   & 0.8788/0.8755    & 0.9002/0.8979  & 0.7183/0.6934 & 0.9553/\underline{\textit{0.9527}}   & 0.5742/0.5552 & 0.8054/0.7949 \\
                        & MUSIQ                  & 0.9646/\underline{\textit{0.9629}}    & 0.9561/0.9437  & 0.8159/0.7647 & 0.7081/0.7344   & 0.7767/0.7697 & 0.8443/0.8351 \\
                        & PAQ2PIQ                & 0.8713/0.8676    & 0.8937/0.8871  & 0.7353/0.7090 & 0.8772/0.7867   & 0.8213/0.7489 & 0.8398/0.7999 \\
                        & WaDIQaM                & 0.9507/0.9430    & 0.9561/0.9522  & 0.7728/\underline{\textit{0.7657}} & 0.7541/0.7510   & 0.7311/0.7512 & 0.8330/0.8326 \\
                        & $MSM_{Unet}$ (ours)         & \underline{\textit{0.9790}}/0.9507    & 0.9723/0.9680  & 0.7996/0.6160 & 0.5696/0.5079   & 0.1980/0.2055 & 0.7037/0.6496 \\
                        & $MSM_{SwinIR}$ (ours)       & 0.9190/0.9043    & 0.8026/0.7722  & 0.7198/0.7349 & 0.7046/0.6700   & 0.7129/0.6932 & 0.7718/0.7549 \\
\SetCell{}{}{}       & $MSM_{Unet}$ (fine-tuned)   & \textbf{0.9810}/\textbf{0.9731}    & \textbf{0.9781}/\textbf{0.9754}  & 0.8275/0.6277 & 0.8745/0.7241   & 0.6685/0.7316 & 0.8659/0.8064 \\
                        & $MSM_{SwinIR}$ (fine-tuned) & 0.9249/0.9024    & 0.8787/0.8233  & 0.6980/0.7256 & 0.6042/0.5525   & 0.6135/0.5641 & 0.7439/0.7136
\end{tblr}
}
\end{table*}

Table 2 compares the evaluation results of six FR-IQA and seven NR-IQA metrics on the T$_1$w MRI Dataset. The averaged results of all five types of distortions are listed in the last column. The highest SRCC and PLCC value is highlighted in \textbf{bold} and the second highest SRCC and PLCC value is marked in\underline{\textit{italics and underline}}. 

In this comparative study, when evaluating images with additional Gaussian noise, Rician noise and DDPM noise, our proposed $MSM_{Unet}$ exhibits a comparative performance to MUSIQ and WaDIQaM, surpassing all other IQA metrics, including FR-IQA metrics. When assessing the images distorted by Motion blur, $MSM_{Unet}$ is surpassed by three FR-IQA metrics, FSIM, LPIPS and DISTS, but still performs better than the listed NR-IQA methods. However, in the evaluation of the Gaussian blurred distortions, two NR-IQA metrics, BRISQUE and NIQE, demonstrate better outcomes than $MSM_{Unet}$. The overall averaged SRCC/PLCC score of the $MSM_{Unet}$ is 0.8705/0.8314, outperforming all listed NR-IQA metrics.

When using SwinIR as the backbone model, $MSM_{SwinIR}$ obtains the most favorable results in assessments of Motion blur, only second to FSIM and DISTS. However, $MSM_{SwinIR}$ does not perform as well as $MSM_{Unet}$ for the images with Gaussian and Rician noise and breast MR images generated from DDPM. The overall averaged SRCC/PLCC score of the $MSM_{SwinIR}$ is 0.8441/0.7926, which is lower than $MSM_{Unet}$, comparable to PAQ2PIQ but outperforming other listed NR-IQA metrics.

Further validation of the MSM method's generalizability was conducted on the Brain Tumor MRI Dataset and generated brain MR images. The results are listed in Table 3, which shows that both $MSM_{Unet}$ and $MSM_{SwinIR}$ maintain comparable performance in noise distortion evaluations across both datasets. $MSM_{Unet}$ outperforms all listed IQA metrics, including FR-IQA metrics. However, assessments of Gaussian and motion blur distortions posed challenges for $MSM_{Unet}$, indicating an inability to effectively gauge these distortions. In contrast, $MSM_{SwinIR}$ sustained moderate performance levels for these blurs, hinting at a more resilient evaluation capacity for structural distortions.

We fine-tuned the breast-trained MSM method for evaluation with the Brain MR images, by using an additional 80 slices of Brain MR images, which are not included in the Brain Tumor MRI Dataset. The fine-tuned results are listed in the last two rows of Table 3. While slight improvements are observed in noise evaluation for both models, $MSM_{Unet}$ significantly benefits in its Gaussian and motion blur evaluation and obtains averaged SRCC/PLCC score of 0.8659/0.8064, outperforming all NR-IQA metrics. Conversely, $MSM_{SwinIR}$ experienced a minor decrease in performance after fine-tuning.

\subsection{Results of Denoised Sodium MR Images }
\begin{table}[ht]
    \centering
    \scriptsize
    \caption{The Cohen’s Kappa results of experts and NR-IQA metrics. *Clinician}
        \centering
        \begin{tabular}{cccc }
        \hline 
        &Expert 1& Expert 2* & Expert 3 \\
        \hline 
        Expert 1  & 1.0000    & 0.4471    & 0.7981  \\ 
        Expert 2* & 0.4471    & 1.0000    & 0.4898  \\ 
        Expert 3  & 0.7981    & 0.4898    & 1.0000   \\ 
        \hline
        
        \end{tabular}
\end{table}
\begin{table}[ht]
    \centering
    \scriptsize
    \caption{The Cohen’s Kappa results of NR-IQA metrics. The highest $\kappa$ scores are highlighted in \textbf{bold}.}
        \centering
        \begin{tabular}{ccccc}
        \hline 
        & Expert 1& Expert 2* & Expert 3 &Average\\
        \hline 
        BRISQUE  & 0.2800    & 0.5200    & 0.2400 &0.3467  \\
        NIQE & 0.0340    & 0.3681    & 0.0097 & 0.1371  \\
        MUSIQ  & 0.5550    & 0.5700    & 0.4057 & 0.5102   \\
        PAQ2PIQ  & 0.6000    & 0.6000    & 0.5600 & 0.5867   \\
       WaDIQaM  & 0.6348    & 0.5700    & 0.5923 & 0.5990 \\
       $MSM_{Unet}$  & 0.5586    & 0.0064    & 0.4373 & 0.3341 \\
       $MSM_{SwinIR}$  & \textbf{0.7173}    & \textbf{0.6468}    & \textbf{0.5942} & \textbf{0.6528} \\
        \hline 
        \end{tabular}
\end{table}

In the process of validating the effectiveness and reliability of the proposed MSM method in evaluating denoised Sodium MR images, we collected the opinions of three experts. Expert 2 is a clinician, Expert 1 and Expert 3 are non-clinical experts in this field. The Cohen’s Kappa coefficient between experts’ assessment results and those of NR-IQA metrics are calculated and listed in Table 4 and Table 5. The best NR-IQA results are highlighted in Bold.

The results suggest the inherent subjectivity in image quality assessment, that individuals may have their preferences in noise level and blurriness. As an example, the $\kappa$ value between Expert 1 and Expert 2 is 0.4471, indicating moderate agreement and diversity in expert opinions. 

$MSM_{Unet}$ exhibits a moderate agreement with Expert 1 and Expert 3 with $\kappa$ values of 0.5586 and 0.4373, which indicates a worse performance than PAQ2PIQ, WaDIQaM and $MSM_{SwinIR}$. Its agreement with Expert 2 is even lower, with an $\kappa$ value of 0.0064, indicating that the agreement is no better than a chance.

On the other hand, despite these variations, the $MSM_{SwinIR}$ evaluation result still has a good alignment with all experts’ assessment results, achieving an average $\kappa$ value of 0.6528, outperforming all other NR-IQA metrics. This suggests that $MSM_{SwinIR}$'s evaluation mechanism is closely aligned with the qualitative judgments of field experts.

These findings support the capability using the MSM method on different DL models, such as the $MSM_{SwinIR}$, to serve as an effective assessment of image quality for image enhancements, such as our denoised sodium MRI images.

\section{Discussion}
In this study, we explored an intrinsic characteristic of deep learning models, defined as Model Specialization, where deviations in the input data from the training data will decrease the model’s prediction performance. The proposed MSM leverages this intrinsic characteristic, avoiding the reliance on external labels.

MSM was initially tested on the T$_1$w MRI Dataset with known levels of noises and distortions. Since MSM was trained exclusively with clean images, it greatly simplifies the data preparation process, requiring only clean images for training. The overall experimental results demonstrated that MSM achieved a remarkable evaluation performance across various types of noises and distortions. Specifically, when evaluating images with additional Gaussian and Rician noises and breast MR images generated from DDPM, $MSM_{Unet}$ exhibited performance comparable to MUSIQ and WaDIQaM, and outperformed all other IQA metrics. While evaluating motion blurriness, $MSM_{SwinIR}$ yielded the most favorable results among all NR-IQA metrics evaluated. Regarding the overall performance across distortions, $MSM_{Unet}$ obtained the highest averaged SRCC/PLCC score among all listed NR-IQA metrics.

Further tests were conducted using the Brain MRI Dataset and generated brain MR images to assess the generalization capabilities of MSM. The findings indicated that both $MSM_{Unet}$ and $MSM_{SwinIR}$ maintained their efficacy in assessing image quality associated with noise. However, when assessing distortions like blurriness, their performance decreased. The performance degradation in $MSM_{SwinIR}$ was considerably less severe compared to $MSM_{Unet}$. Notably, via fine-tuning, the model’s performance, especially $MSM_{Unet}$, was greatly enhanced. It suggests a potential weakness to overfitting due to the unique training strategy of backbone models and limited volume of the training dataset. Diversifying the training dataset or model fine-tuning addresses this limitation. This result also indicates that if the clean images for training have similar structures as the testing images, it will improve the evaluation performance of proposed MSM.

The performance variance between $MSM_{Unet}$ and $MSM_{SwinIR}$ can be attributed to the intrinsic differences in feature extraction capabilities of CNN-based models versus Transformer-based models. CNN-based models use convolution kernels with shared weights to extract local features, which offers some position invariance but can struggle with identifying structural artefacts requiring long-range dependencies. Conversely, Transformer-based models are designed to capture global spatial relationships, making them advantageous for tasks where spatial relationships are crucial. It is important to note that while transformer-based models excel in capturing global dependencies, CNN-based models are typically better at extracting local features and perform well on small to medium-sized datasets. The results emphasized the importance of the backbone architecture's role in model performance. $MSM_{Unet}$ has an excellent evaluation performance in noise-related distortions assessments, where image structures play a less critical role. Conversely, when the distortion is associated with the image structures, such as blurriness, $MSM_{SwinIR}$ exhibits a better evaluation performance than $MSM_{Unet}$.  

Subsequent validation with denoised sodium MR images reaffirmed the effectiveness of $MSM_{SwinIR}$, demonstrating remarkable alignment with experts’ evaluations, achieving the highest averaged Cohen’s Kappa coefficient of 0.6528, outperforming other NR-IQA metrics. Therefore, the proposed MSM provides a reliable and label-free tool for assessing the quality of denoised authentic sodium MR images where no high-quality reference exists.

For further research, a meaningful approach involves the development of a hybrid backbone model that integrates the strengths of both CNN-based and Transformer-based architectures. By integrating CNN-based models' local feature extraction capabilities with Transformer-based models' structural sensitivity, there is potential to create a more versatile and effective tool for image quality assessment.

\section{Conclusion}
This study confronts the challenge inherent in assessing the quality of sodium MRI. Sodium MRI lacks high-quality reference images, which impedes the effectiveness of evaluating enhanced images with conventional FR-IQA metrics. Addressing this critical issue, we introduce a DL-based, label-free NR-IQA metric, called Model Specialization Metric. This innovative approach provides new insight into the application of DL models for IQA tasks and has shown consistency with expert evaluations with an average Cohen’s Kappa coefficient of 0.6528, outperforming existing NR-IQA metrics. The capability for evaluating referenceless image quality is particularly valuable in medical imaging, where the availability of such baseline datasets can be limited or non-existent.

\section*{Acknowledgment}
We acknowledge support from Cancer Research UK (A25922), the Medical Research Council (MR/X018067/1) and CRUK Cambridge Centre and the NIHR Cambridge Biomedical Research Centre (BRC-121520014). This work was performed using resources provided by the Cambridge Service for Data Driven Discovery (CSD3) operated by the University of Cambridge Research Computing Service (www.csd3.cam.ac.uk), provided by Dell EMC and Intel using Tier-2 funding from the Engineering and Physical Sciences Research Council (capital grant EP/T022159/1), and DiRAC funding from the Science and Technology Facilities Council (www.dirac.ac.uk).



\begin{thebibliography}{00} 
\bibitem{b1} R. Ouwerkerk et al., “Elevated tissue sodium concentration in malignant breast lesions detected with non-invasive 23Na MRI,” Breast Cancer Res Treat, vol. 106, no. 2, pp. 151–160, Dec. 2007, doi: 10.1007/s10549-006-9485-4.

\bibitem{b2} E. Murphy and D. A. Eisner, “Regulation of Intracellular and Mitochondrial Sodium in Health and Disease,” Feb. 13, 2009. doi: 10.1161/CIRCRESAHA.108.189050.

\bibitem{b3} G. Madelin and R. R. Regatte, “Biomedical applications of sodium MRI in vivo,” \textit{Journal of Magnetic Resonance Imaging}, vol. 38, no. 3, pp. 511–529, Sep. 2013, doi: 10.1002/jmri.24168.

\bibitem{b4} G. Madelin, J. S. Lee, R. R. Regatte, and A. Jerschow, “Sodium MRI: Methods and applications,” 2014, \textit{Elsevier B.V.} doi: 10.1016/j.pnmrs.2014.02.001. 

\bibitem{b5} S. Nielles-Vallespin \textit{et al.}, “3D radial projection technique with ultrashort echo times for sodium MRI: Clinical applications in human brain and skeletal muscle,” \textit{Magn Reson Med}, vol. 57, no. 1, pp. 74–81, 2007, doi: 10.1002/mrm.21104. 

\bibitem{b6} H. Schomberg and J. Timmer, “The Gridding Method for Image Reconstruction by Fourier Transformation,” 1995. 

\bibitem{b7} J. A. Fessler, “On NUFFT-based gridding for non-Cartesian MRI,” \textit{Journal of Magnetic Resonance}, vol. 188, no. 2, pp. 191–195, Oct. 2007, doi: 10.1016/j.jmr.2007.06.012. 

\bibitem{b8} T. Knopp, S. Kunis, and D. Potts, “A note on the iterative MRI reconstruction from nonuniform k-space data,” \textit{Int J Biomed Imaging}, vol. 2007, 2007, doi: 10.1155/2007/24727.

\bibitem{b9} S. Kaji and S. Kida, “Overview of image-to-image translation by use of deep neural networks: denoising, super-resolution, modality conversion, and reconstruction in medical imaging,” Sep. 01, 2019, \textit{Springer Tokyo}. doi: 10.1007/s12194-019-00520-y. 

\bibitem{b10} H. Yang, Z. Wang, X. Liu, C. Li, J. Xin, and Z. Wang, “Deep learning in medical image super resolution: a review,” \textit{Applied Intelligence}, vol. 53, no. 18, pp. 20891–20916, Sep. 2023, doi: 10.1007/s10489-023-04566-9.

\bibitem{b11} Z. Wang, A. C. Bovik, H. R. Sheikh, and E. P. Simoncelli, “Image quality assessment: From error visibility to structural similarity,” \textit{IEEE Transactions on Image Processing}, vol. 13, no. 4, pp. 600–612, Apr. 2004, doi: 10.1109/TIP.2003.819861.

\bibitem{b12} L. Zhang, L. Zhang, X. Mou, and D. Zhang, “FSIM: A feature similarity index for image quality assessment,” \textit{IEEE Transactions on Image Processing}, vol. 20, no. 8, pp. 2378–2386, Aug. 2011, doi: 10.1109/TIP.2011.2109730. 

\bibitem{b13} A. Mittal, A. K. Moorthy, and A. C. Bovik, “No-reference image quality assessment in the spatial domain,” \textit{IEEE Transactions on Image Processing}, vol. 21, no. 12, pp. 4695–4708, 2012, doi: 10.1109/TIP.2012.2214050.

\bibitem{b14} I. Stępień and M. Oszust, “A Brief Survey on No-Reference Image Quality Assessment Methods for Magnetic Resonance Images,” \textit{J Imaging}, vol. 8, no. 6, Jun. 2022, doi: 10.3390/jimaging8060160. 

\bibitem{b15} K. Simonyan and A. Zisserman, “Very Deep Convolutional Networks for Large-Scale Image Recognition,” Sep. 2014, [Online]. Available: http://arxiv.org/abs/1409.1556 

\bibitem{b16} K. He, X. Zhang, S. Ren, and J. Sun, “Deep Residual Learning for Image Recognition.” [Online]. Available: http://image-net.org/challenges/LSVRC/2015/ 

\bibitem{b17} A. Dosovitskiy \textit{et al.}, “An Image is Worth 16x16 Words: Transformers for Image Recognition at Scale,” Oct. 2020, [Online]. Available: http://arxiv.org/abs/2010.11929

\bibitem{b18} M. Oszust, M. Bielecka, A. Bielecki, I. Ste¸pień, R. Obuchowicz, and A. Piórkowski, “Blind image quality assessment of magnetic resonance images with statistics of local intensity extrema,” \textit{Inf Sci (N Y)}, vol. 606, pp. 112–125, Aug. 2022, doi: 10.1016/j.ins.2022.05.061. 

\bibitem{b19} Q. Chen \textit{et al.}, “MRIQA: Subjective Method and Objective Model for Magnetic Resonance Image Quality Assessment,” in \textit{2022 IEEE International Conference on Visual Communications and Image Processing, VCIP 2022}, Institute of Electrical and Electronics Engineers Inc., 2022. doi: 10.1109/VCIP56404.2022.10008885.

\bibitem{b20} S. Yu, G. Dai, Z. Wang, L. Li, X. Wei, and Y. Xie, “A consistency evaluation of signal-to-noise ratio in the quality assessment of human brain magnetic resonance images,” \textit{BMC Med Imaging}, vol. 18, no. 1, May 2018, doi: 10.1186/s12880-018-0256-6.

\bibitem{b21} J. Kim, A. D. Nguyen, and S. Lee, “Deep CNN-Based Blind Image Quality Predictor,” \textit{IEEE Trans Neural Netw Learn Syst}, vol. 30, no. 1, pp. 11–24, Jan. 2019, doi: 10.1109/TNNLS.2018.2829819.

\bibitem{b22} P. Ye, J. Kumar, and D. Doermann, “Beyond Human Opinion Scores: Blind Image Quality Assessment based on Synthetic Scores.”

\bibitem{b23} M. S. Treder, R. Codrai, and K. A. Tsvetanov, “Quality assessment of anatomical MRI images from generative adversarial networks: Human assessment and image quality metrics,” \textit{J Neurosci Methods}, vol. 374, May 2022, doi: 10.1016/j.jneumeth.2022.109579.

\bibitem{b24} R. C. Streijl, S. Winkler, and D. S. Hands, “Mean opinion score (MOS) revisited: methods and applications, limitations and alternatives,” \textit{Multimed Syst}, vol. 22, no. 2, pp. 213–227, Mar. 2016, doi: 10.1007/s00530-014-0446-1.

\bibitem{b25} O. Ronneberger, P. Fischer, and T. Brox, “U-net: Convolutional networks for biomedical image segmentation,” in \textit{Lecture Notes in Computer Science (including subseries Lecture Notes in Artificial Intelligence and Lecture Notes in Bioinformatics)}, Springer Verlag, 2015, pp. 234–241. doi: 10.1007/978-3-319-24574-4\_28. 

\bibitem{b26} N. Ponomarenko \textit{et al.}, “Image database TID2013: Peculiarities, results and perspectives,” \textit{Signal Process Image Commun}, vol. 30, pp. 57–77, Jan. 2015, doi: 10.1016/j.image.2014.10.009. 

\bibitem{b27} J. Liang, J. Cao, G. Sun, K. Zhang, L. Van Gool, and R. Timofte, “SwinIR: Image Restoration Using Swin Transformer.” [Online]. Available: https://github.com/JingyunLiang/SwinIR

\bibitem{b28} K. Krupa and M. Bekiesińska-Figatowska, “Artifacts in magnetic resonance imaging,” \textit{Pol J Radiol}, vol. 80, no. 1, pp. 93–106, Feb. 2015, doi: 10.12659/PJR.892628.

\bibitem{b29} G. Bradski, “The OpenCV Library,” \textit{Dr. Dobb’s Journal of Software Tools}, 2000. 

\bibitem{b30} J. Ho, A. Jain, and P. Abbeel, “Denoising Diffusion Probabilistic Models.” [Online]. Available: https://github.com/hojonathanho/diffusion. 

\bibitem{b31} K. Dabov, A. Foi, V. Katkovnik, and K. Egiazarian, “Image denoising by sparse 3-D transform-domain collaborative filtering,” \textit{IEEE Transactions on Image Processing}, vol. 16, no. 8, pp. 2080–2095, Aug. 2007, doi: 10.1109/TIP.2007.901238. 

\bibitem{b32} K. Zhang, W. Zuo, Y. Chen, D. Meng, and L. Zhang, “Beyond a Gaussian denoiser: Residual learning of deep CNN for image denoising,” \textit{IEEE Transactions on Image Processing}, vol. 26, no. 7, pp. 3142–3155, Jul. 2017, doi: 10.1109/TIP.2017.2662206. 

\bibitem{b33} Z. Zhang, Q. Liu, and Y. Wang, “Road Extraction by Deep Residual U-Net,” \textit{IEEE Geoscience and Remote Sensing Letters}, vol. 15, no. 5, pp. 749–753, May 2018, doi: 10.1109/LGRS.2018.2802944.

\bibitem{b34} S. M. Vieira, U. Kaymak, and J. M. C. Sousa, “Cohen’s kappa coefficient as a performance measure for feature selection,” in \textit{2010 IEEE World Congress on Computational Intelligence, WCCI 2010}, 2010. doi: 10.1109/FUZZY.2010.5584447.

\bibitem{b35} J. Kaggie \textit{et al.}, “Sodium Breast Imaging of Ductal Carcinomas at 3 T,” doi: 10.58530/2022/1684.

\bibitem{b36} Artem Shlezinger, “Brain Tumor MRI,” https://www.kaggle.com/datasets/shlezinger/brain-mri-data/data. 

\bibitem{b37} R. Zhang, P. Isola, A. A. Efros, E. Shechtman, and O. Wang, “The Unreasonable Effectiveness of Deep Features as a Perceptual Metric.” [Online]. Available: https://www.github.com/richzhang/PerceptualSimilarity.

\bibitem{b38} K. Ding, K. Ma, S. Wang, and E. P. Simoncelli, “Image Quality Assessment: Unifying Structure and Texture Similarity,” \textit{IEEE Trans Pattern Anal Mach Intell}, vol. 44, no. 5, pp. 2567–2581, May 2022, doi: 10.1109/TPAMI.2020.3045810. 

\bibitem{b39} A. Mittal, R. Soundararajan, and A. C. Bovik, “Making a ‘completely blind’ image quality analyzer,” \textit{IEEE Signal Process Lett}, vol. 20, no. 3, pp. 209–212, 2013, doi: 10.1109/LSP.2012.2227726. 

\bibitem{b40} Z. Ying, H. Niu, P. Gupta, D. Mahajan, D. Ghadiyaram, and A. Bovik, “From Patches to Pictures (PaQ-2-PiQ): Mapping the Perceptual Space of Picture Quality.”

\bibitem{b41} J. Ke, Q. Wang, Y. Wang, P. Milanfar, and F. Yang, “MUSIQ: Multi-scale Image Quality Transformer.” [Online]. Available: https://github.com/ 

\bibitem{b42} S. Bosse, D. Maniry, K. R. Müller, T. Wiegand, and W. Samek, “Deep Neural Networks for No-Reference and Full-Reference Image Quality Assessment,” \textit{IEEE Transactions on Image Processing}, vol. 27, no. 1, pp. 206–219, Jan. 2018, doi: 10.1109/TIP.2017.2760518. 

\bibitem{b43} C. Chen and J. Mo, “IQA-PyTorch: Pytorch toolbox for image quality assessment,” https://github.com/chaofengc/IQA-PyTorch. 

\bibitem{b44} H. Lin, V. Hosu, and D. Saupe, “KADID-10k: A large-scale artificially distorted IQA database,” in \textit{2019 Eleventh International Conference on Quality of Multimedia Experience (QoMEX)}, Mar. 2019. doi: 10.1109/TIP.2020.2967829.

\bibitem{b45} V. Hosu, H. Lin, T. Sziranyi, and D. Saupe, “KonIQ-10k: An Ecologically Valid Database for Deep Learning of Blind Image Quality Assessment,” \textit{IEEE Transactions on Image Processing}, vol. 29, pp. 4041–4056, 2020, doi: 10.1109/TIP.2020.2967829. 

\bibitem{b46} J. Johnson, Alexandre Alahi, and Li Fei-Fei, “Perceptual losses for real-time style transfer and super-resolution,” in \textit{Computer Vision–ECCV 2016: 14th European Conference}, B. Leibe, J. Matas, N. Sebe, and M. Welling, Eds., in Lecture Notes in Computer Science, vol. 9906. Cham: Springer International Publishing, 2016, pp. 694–711. doi: 10.1007/978-3-319-46475-6.  
\end{thebibliography}
\end{document}